\documentclass{iucrjournals}

\usepackage{amsfonts}
\usepackage{amsmath}
\usepackage{graphicx} 
\usepackage{float}    
\usepackage{subcaption} 

\title{Machine Learning Derived Blood Input for Dynamic PET Images of Rat Heart}

\author[a,b]{Shubhrangshu Debsarkar\IUCrCemaillink{fvc9ch@virignia.edu}\IUCrOrcidlink{xxxx-xxxx-xxxx-xxxx}}%
\author[b,c]{Bijoy Kundu\IUCrEmaillink{bkk5a@virginia.edu}\IUCrOrcidlink{xxxx-xxxx-xxxx-xxxx}}%

\affil[a]{Computer Science, University of Virginia , Charlottesville, VA, United States }
\affil[b]{Radiology, University of Virginia, Charlottesville, VA, United States }
\affil[c]{Biomedical Engineering, Charlottesville, VA, United States}

\begin{document} 
\nolinenumbers
\maketitle 

\begin{abstract}
Dynamic FDG PET imaging study of $n=52$ rats including 26 control Wistar-Kyoto (WKY) rats and 26 experimental spontaneously hypertensive rats (SHR) were performed using a Siemens microPET and Albira trimodal scanner longitudinally at 1, 2, 3, 5, 9, 12 and 18 months of age. A 15-parameter dual output model correcting for spill over contamination and partial volume effects with peak fitting cost functions was developed for simultaneous estimation of model corrected blood input function (MCIF) and kinetic rate constants for dynamic FDG PET images of rat heart in vivo. Major drawbacks of this model are its dependence on manual annotations for the Image Derived Input Function (IDIF) and manual determination of crucial model parameters to compute MCIF. To overcome these limitations, we performed semi-automated segmentation and then formulated a Long-Short-Term Memory (LSTM) cell network to train and predict MCIF in test data using a concatenation of IDIFs and myocardial inputs and compared them with reference-modeled MCIF. Thresholding along 2D plane slices with two thresholds, with T1 representing high-intensity myocardium, and T2 representing lower-intensity rings, was used to segment the area of the LV blood pool. The resultant IDIF and myocardial TACs were used to compute the corresponding reference (model) MCIF for all data sets. The segmented IDIF and the myocardium formed the input for the LSTM network.  A k-fold cross validation structure with a 33:8:11 split and 5 folds was utilized to create the model and evaluate the performance of the LSTM network for all datasets. To overcome the sparseness of data as time steps increase, midpoint interpolation was utilized to increase the density of datapoints beyond time = 10 minutes. The model utilizing midpoint interpolation was able to achieve a $56.4\%$ improvement over previous Mean Squared Error (MSE). 
\end{abstract}


\section{Introduction}
\subsection{PET}
Positron emission tomography (PET) is a non-invasive molecular imaging technique that has been widely utilized in clinical and research settings for several decades \cite{phelps1975application}. Unlike anatomical imaging modalities such as magnetic resonance imaging (MRI) and computed tomography (CT), which primarily provide high-resolution structural details, PET allows for in vivo visualization of physiological and biochemical processes within the body. The fundamental mechanism underlying PET imaging involves the use of radiotracers—biologically active molecules labeled with positron-emitting radionuclides. Upon administration, these radiotracers accumulate in specific tissues based on metabolic activity and are subsequently detected by PET scanners, enabling the identification of regions with altered biochemical function.

One of the most commonly used radiotracers in PET imaging is 2-[18F] fluoro-2-deoxy-D-glucose (FDG), a glucose analog that provides insight into glucose metabolism at the cellular level. Since malignant cells typically exhibit increased glucose uptake due to heightened metabolic activity, FDG-PET has become a critical tool in oncology for tumor detection, staging, and treatment response assessment. In addition to oncology, PET imaging has broad applications in neurology and cardiology. In neuroimaging, FDG-PET is extensively used to assess metabolic dysfunction in neurodegenerative diseases, such as Alzheimer’s disease and other dementias, by identifying regions of hypometabolism associated with neuronal degeneration. Furthermore, PET plays a significant role in epilepsy evaluation, where it aids in identifying epileptogenic foci in patients with medically refractory epilepsy. 

Beyond FDG, a range of other PET radiotracers have been developed to target different biological pathways, expanding the utility of PET imaging. For instance, oxygen-15-labeled water is frequently used in cardiac PET studies to assess myocardial perfusion, while amyloid- and tau-specific PET tracers provide critical information in Alzheimer’s disease research. The combination of PET with CT or MRI (PET/CT, PET/MRI) has further enhanced diagnostic capabilities by integrating functional and structural imaging, allowing for more precise disease localization and characterization.
\subsection{dPET}
Static FDG-PET is the most prevalent modality of scanning. Patients are injected with radiotracer and typically imaged 45 minutes after injection. The resultant snapshot is typically only qualitatively useful and does not provide quantitively analyzable information. On the other hand, Dynamic FDG-PET allows for a greater quantification of data. Patients are put on a scanner for 60-90 minutes and brain activity is imaged at discrete time thorughout the scan. The resultant data is a time-series of 3 Dimensional Scans. dFDG-PET provides clinicians with much more information on \textbf{in vivo} pathway activity. 
\subsection{IDIF}
Quantitative analysis of dynamic PET scans is based on the blood input function, a time-activity curve (TAC), which tracks the concentration of radiotracers available for the tissue to use at each time step. The gold standard for measuring this function, arterial blood sampling, is invasive, costly, and carries risks such as infection and arterial occlusion. In the case of rodents specifically, small size in addition to a high metabolism rate makes arterial blood sampling incredibly difficult \cite{laforest2005measurement}. To overcome these challenges, researchers have developed the image-derived blood input function (IDIF), which non-invasively estimates radiotracer kinetics from PET image data. For humans, the carotid artery is used as the region of interest. For this study, we will be analyzing intensities in two areas: The left ventricle blood pool and the surrounding myocardial tissue. 

\begin{figure}[H] 
    \centering
    \includegraphics[width=\textwidth]{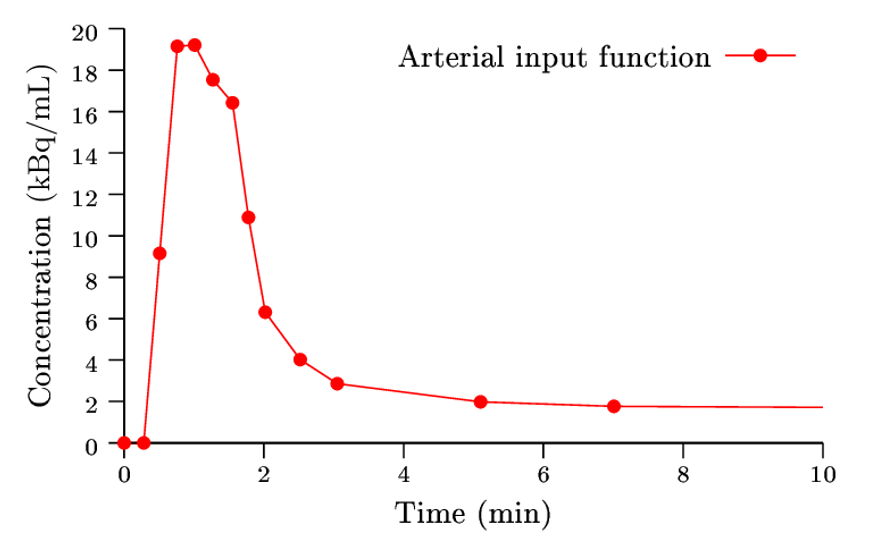}
    \caption{A typical input function for a specific rodent. Each point represents a short scan of the body. Notice that the intervals between scans are not uniformly distributed. Only the first ten minutes are shown for representation.} 
    \label{fig:input_function} 
\end{figure}

\begin{figure}[H] 
    \centering
    \includegraphics[width=\textwidth]{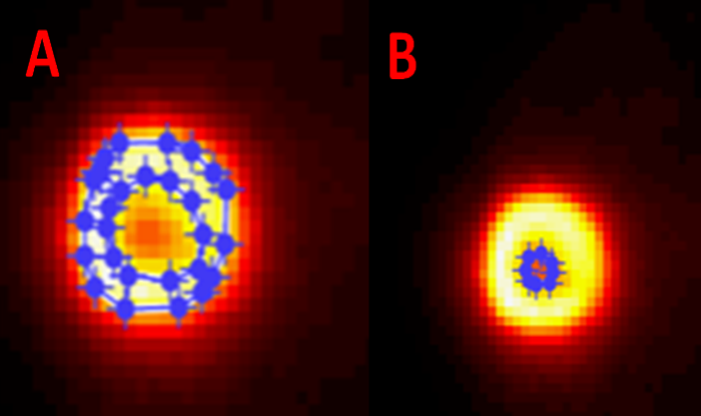}
    \caption{2D map showing segmentation points (blue) of a rodent heart. Image A shows the segmentation of the myocardial tissue. Image B shows the segmentation of the left ventricule blood pool. } 
    \label{fig:segmentation_diagram} 
\end{figure}

Despite its benefits, IDIF adoption is hindered by manual carotid artery segmentation, a laborious and imprecise process that requires expert annotation. To address this limitation, this study employs supervised machine learning (ML) to automate carotid artery segmentation and IDIF computation. As can be seen in Figure \ref{fig:segmentation_diagram}, segmentations allow researchers to focus on the most important parts of the PET scan, which are the LV pool and surrounding myocardial tissue in the case of rodents. 

\subsection{MCIF}
PET’s inherently low spatial resolution introduces challenges in accurately estimating the image-derived blood input function (IDIF). Two major sources of error are spillover (SP) and partial volume (PV) effects. SP occurs when signal from high-activity regions contaminates adjacent lower-activity regions, leading to overestimated tracer concentration, while PV effects arise from voxel averaging, causing underestimation in small structures like the carotid arteries. These issues compromise IDIF accuracy, particularly in regions smaller than the scanner’s spatial resolution. 

To correct for these effects, we apply a three-compartment model to derive the Model-Corrected Input Function (MCIF), refining IDIF estimates by incorporating surrounding tissue activity \cite{massey2021model}. Comping the MCIF requires optimizing two objective functions, one to match general curve shapes and one to align the peaks. However, the model’s complexity (15 parameters) and non-convex optimization required manual parameter bounding, limiting scalability. This model has been validated against ground truth data collected through arterial sampling.

To overcome these constraints, this study introduces a deep learning-based approach for MCIF estimation. Using a recurrent neural network (RNN) architecture, we aim to automate parameter optimization and predict MCIF directly from IDIF, improving accuracy while reducing manual intervention \cite{sun2018predicting}. This approach enhances blood input function estimation, addressing PET’s resolution limitations and facilitating more reliable quantification in dynamic PET imaging.


\section{Datasets}

\subsection{Rodent Imaging Datasets}
Dynamic FDG-PET imaging was performed on 52 rats (26 control Wistar-Kyoto [WKY] and 26 spontaneously hypertensive rats [SHR]) using Siemens microPET and Albira trimodal scanners with longitudinal scans at 1, 2, 3, 5, 9, 12, and 18 months of age \cite{li2019metabolic}. The rationale for the usage of variegated ages and scanners was to make the model generalizable to rodents of all types \cite{li2023progressive}. Time frames were binned with higher temporal resolution during early phases (8-second intervals initially) to capture rapid tracer dynamics and first-pass kinetics, progressing to 6-minute frames during metabolic equilibrium phases. Each scan spanned 60 minutes with variable temporal sampling density, prioritizing early timepoints to resolve peak blood pool activity. The reconstructed PET data exhibited spatial resolutions consistent with scanner specifications (Siemens microPET: $ < 2$ mm³ voxels; Albira:  $< 1.5$ mm³ voxels). Semi-automated segmentation of left ventricular (LV) blood pool and myocardial tissue was performed using dual-threshold analysis (T1 = high-intensity myocardium, T2 = low-intensity blood pool) across all timeframes.

\subsection{LSTM Training Datasets}
The neural network input comprised $\mathbb{R}^{N \times 23 \times 2}$ tensors containing normalized IDIFs from the LV blood pool and myocardial TACs at 23 time points \cite{schmidhuber2002learning}. The sequences contained 23 timepoints with increasing sparseness of points as time increased.  All TACs were intensity-normalized by their corresponding IDIF peak values. The dataset included 364 total scans (52 rodents $\times$ 7 longitudinal timepoints) partitioned into 33:8:11 train:validation:test splits with 5-fold cross-validation. This partitioning ensured balanced representation across scanner types (microPET/Albira), rodent strains (WKY/SHR), and age cohorts (1-18 months). Ground truth MCIFs were derived using a validated 15-parameter dual-output model with spillover correction and partial volume recovery, serving as regression targets for the LSTM network \cite{li2018improved} \cite{huang2019non}.

\section{Interpolation}

\subsection{Rationale for Interpolation}
Dynamic FDG-PET imaging data exhibits high temporal resolution during the initial phases of acquisition, with frames collected at 8-second intervals to capture rapid tracer dynamics and peak blood input activity. However, as the scan progresses, temporal sampling becomes increasingly sparse, with frames collected every 6 minutes after approximately 10 minutes. This progressive sparsity reduces the ability to accurately characterize the tail behavior of time-activity curves (TACs), which is critical for deriving accurate blood input functions. To address this limitation, midpoint interpolation was employed to artificially increase the temporal resolution of TACs in later phases, ensuring uniform sequence lengths and improving model performance.

\subsection{Interpolation Methodology}
Midpoint interpolation was applied to all TACs (image-derived input function [IDIF] and myocardial tissue TACs) beyond $t = 10$ minutes. For each pair of consecutive time points $t_i$ and $t_{i+1}$, a new interpolated point $C_{\text{interp}}(t)$ was calculated as:
\[
C_{\text{interp}}(t) = \frac{C(t_i) + C(t_{i+1})}{2}, \quad t = \frac{t_i + t_{i+1}}{2}.
\]
This method effectively increased the sequence length from 23 to 30 time points, filling in gaps in the later phases of the TACs while preserving the physiological trends observed in the original data.

\subsection{Impact of Interpolation on Data Characteristics}
Before interpolation, TACs contained 23 time points with irregular temporal resolution, particularly sparse beyond $t = 10$ minutes. After interpolation, all TACs were standardized to 30 time points with uniform spacing in later phases. This augmentation improved the ability of the Long Short-Term Memory (LSTM) network to predict model-corrected input functions (MCIFs) by providing a more complete representation of tracer kinetics throughout the scan duration.
\begin{table}[h!]
\centering
\caption{Comparison of Non-Interpolated and Interpolated Datasets}
\renewcommand{\arraystretch}{1.2} 
\setlength{\tabcolsep}{12pt} 
\begin{tabular}{|c|c|c|}
\hline
\textbf{Property} & \textbf{Non-Interpolated} & \textbf{Interpolated} \\ \hline
Number of Time Points & 23 & 30 \\ \hline
Temporal Resolution (Early Phase) & 8 seconds & 8 seconds \\ \hline
Temporal Resolution (Late Phase) & 6 minutes & 3 minutes \\ \hline
Average Distance Between Points (Late Phase) & 6 minutes & 3 minutes \\ \hline
\end{tabular}
\label{tab:dataset_comparison}
\end{table}

\subsection{Effect on Model Performance}
The inclusion of interpolated data allows for the predicted MCIF curve to very closely match the actual MCIF curve on the tails. This can lead to a significant improvement in performance on metrics such as mean squared error and aid in various downstream tasks.

\begin{figure}[H] 
    \centering
    \includegraphics[width=\textwidth]{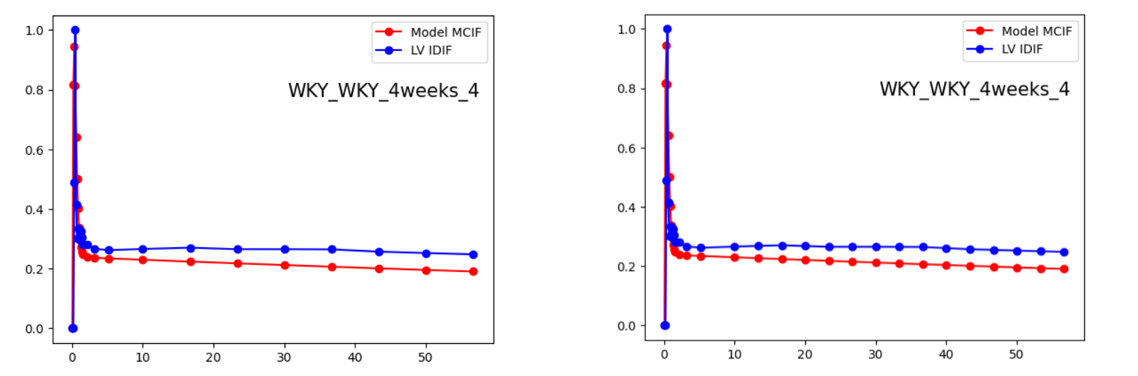}
    \caption{2 graphs of IDIF and model MCIF curves for an example rodent. Left: base curves with no interpolation. Right: Curves with interpolation after t = 10 minutes.} 
    \label{fig:interp_vs_nointerp} 
\end{figure}

\section{MCIF Architecture}

The task of mapping image-derived input functions (IDIFs) and myocardial time-activity curves (TACs) to model-corrected input functions (MCIFs) was addressed using a Long Short-Term Memory (LSTM) neural network. The LSTM network was selected due to its ability to model temporal dependencies effectively, making it well-suited for time-series regression tasks such as this.

\subsection*{Model Inputs and Outputs}
The input to the model consisted of a tensor of dimensions $\mathbb{R}^{N \times 30 \times 2}$, where $N$ represents the number of samples, 30 corresponds to the interpolated time points, and 2 represents the dual-channel input comprising normalized IDIFs and myocardial TACs. The IDIF and myocardial TACs were intensity-normalized by dividing each curve by the peak value of the corresponding IDIF. The target output was a tensor of dimensions $\mathbb{R}^{N \times 30 \times 1}$, representing the predicted MCIF values at each interpolated time point.

\subsection*{Network Architecture}
The architecture of the LSTM network was designed to balance computational efficiency and predictive accuracy. It consisted of a single LSTM layer with 1000 units, followed by a time-distributed dense layer with a single output unit \cite{sherstinsky2020fundamentals}. The LSTM layer included a tanh activation function, which is standard for capturing non-linear temporal relationships in sequential data. The dense layer applied a linear activation function to produce continuous-valued outputs corresponding to MCIF predictions.

Mathematically, the LSTM layer can be represented as follows:
\[
h_t = \sigma(W_h x_t + U_h h_{t-1} + b_h),
\]
where $h_t$ is the hidden state at time $t$, $x_t$ is the input at time $t$, $W_h$ and $U_h$ are weight matrices, $b_h$ is the bias vector, and $\sigma$ represents the activation function.

The time-distributed dense layer maps each hidden state $h_t$ to an output $y_t$:
\[
y_t = W_o h_t + b_o,
\]
where $W_o$ is the weight matrix and $b_o$ is the bias vector.
\begin{figure}[H] 
    \centering
    \includegraphics[width=\textwidth]{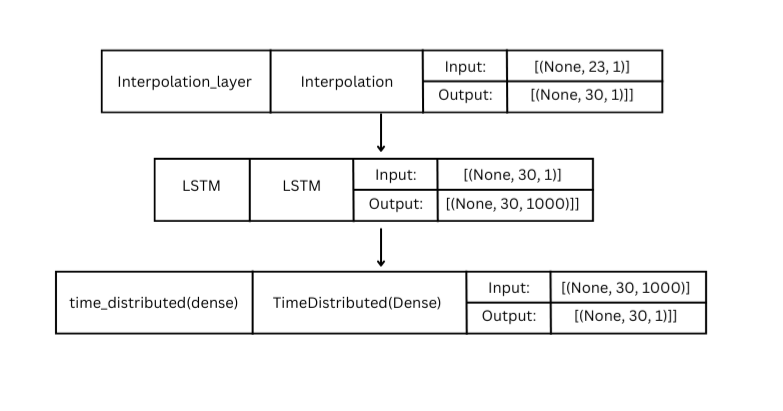}
    \caption{The model architecture used to predict MCIFs from a concatenation of IDIFs and myocardial inputs - a single layer of 1000 LSTM cells, then time-distributed over a dense layer to regress the predicted MCIF using an interpolated 30 frame time series as input. } 
    \label{fig:model_architecture} 
\end{figure}

\subsection*{Training Configuration}
The model was trained using the Adam optimizer with a learning rate of $0.0001$. The loss function employed was mean squared error (MSE), defined as:
\[
\mathcal{L}_{MSE} = \frac{1}{N} \sum_{i=1}^{N} (y_{\text{true}}^{(i)} - y_{\text{pred}}^{(i)})^2,
\]
where $y_{\text{true}}^{(i)}$ and $y_{\text{pred}}^{(i)}$ represent the ground truth and predicted MCIF values for sample $i$, respectively. Early stopping was implemented with a patience of 5 epochs and a minimum delta of 0.001 to prevent overfitting. MSE was chosen due to its effectiveness in matching each point from the IDF to the MCIF, allowing both curves to coincide as much as possible without reducing matches on important parts of the curve such as the peak. 

The dataset was split into training, validation, and test sets in a ratio of 33:8:11 using a k-fold cross-validation approach with 5 folds. Each fold ensured that data from individual rodents did not appear in both training and validation sets simultaneously.

The metrics used for validation were MSE as described above and Dynamic Time Warping (DTW) distance. DTW is a distance metric used to measure the similarity between two temporal sequences that may vary in speed or length. Given two sequences
\[
A = (a_1, a_2, \dots, a_n), \quad B = (b_1, b_2, \dots, b_m),
\]
DTW finds the optimal alignment by minimizing the cumulative cost of matching points from $A$ to $B$ through warping in the time dimension.

The cost matrix $D \in \mathbb{R}^{n \times m}$ is defined by the squared Euclidean distance:
\[
D(i, j) = (a_i - b_j)^2.
\]
The cumulative cost matrix $C$ is computed recursively as:
\[
C(i, j) = D(i, j) + \min\left\{
\begin{array}{l}
C(i-1, j), \\
C(i, j-1), \\
C(i-1, j-1)
\end{array}
\right.,
\]
with appropriate boundary conditions. The DTW distance is given by:
\[
\text{DTW}(A, B) = C(n, m).
\]

\subsection*{Hyperparameters and Implementation Details}
The total number of trainable parameters in the model was 4,013,001. Training was conducted over a maximum of 1000 epochs with a batch size of 32; however, training typically converged between 350–500 epochs due to early stopping. The implementation utilized TensorFlow as the deep learning framework, and all computations were performed on an NVIDIA RTX 4080 GPU with 16 GB VRAM.

This architecture demonstrated robust performance in predicting MCIFs from IDIFs and myocardial TACs, leveraging temporal dependencies effectively while maintaining computational efficiency.

\begin{figure}[H] 
    \centering
    \includegraphics[width=\textwidth]{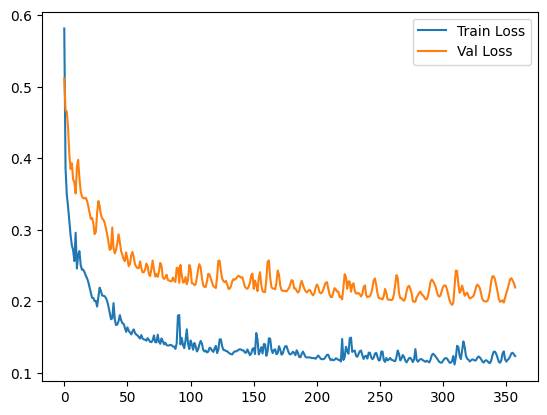}
    \caption{Loss curve for the convergence of the algorithm. The x-axis represents epochs and the y-axis represents MSE. Early stopping can be seen in the fact that the model cuts off after $\sim$375 epochs.} 
    \label{fig:loss_curve} 
\end{figure}

\section{Results}
Results of our interpolated and non-interpolated LSTM were evaluated based on DTW and MSE metrics as discussed earlier in the methods section. 
\subsection{Non-interpolated LSTM Results}
The base, non-interpolated model was able to achieve an MSE of $0.1176 \pm 0.0132$ and a DTW of $6.1478 \pm 0.3702$. This is consistent with previous methodologies that attempted to use similar techniques to map MCIFs \cite{qureshi2022machine}. These results and error bars are the averages and standard deviations across all five folds.  

\subsection{Interpolated LSTM Results}
The interpolated model was able to achieve an MSE of $0.0806 \pm 0.0107$ and a DTW of $6.0898 \pm 0.4342$. This is a strict increase from the non-interpolated results from this paper as well as results from previous investigations. The effectiveness of interpolation can clearly be seen in many of the graphs comparing the ground truth MCIF and the predicted MCIF with interpolation, as shown below. 

\begin{figure}[H] 
    \centering
    \includegraphics[width=\textwidth]{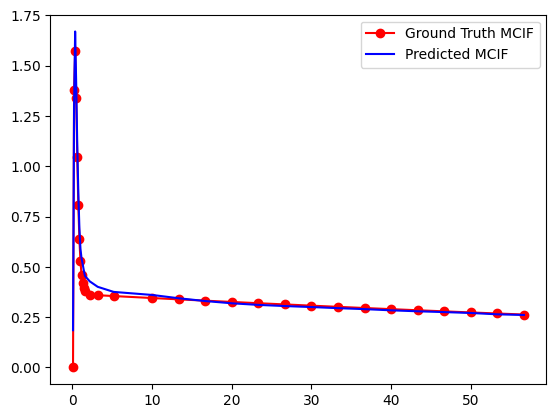}
    \caption{A graph showing the ground truth MCIF curve (red) compared to the predicted TAC by the interpolated model (blue). The x-axis is the time in minutes and the y-axis is the activity concentration.} 
    \label{fig:loss_curve} 
\end{figure}

\subsection{Comparison}
We observe a statistically significant difference between the mean squared errors of the two methods, suggesting an LSTM trained on an interpolated dataset will be more effective at closely matching both the tail of the MCIF and its peak. The interpolated model achieved an improvement of $56.4\%$ in MSE compared to previous methods \cite{qureshi2022machine}. Although it introduced another pre-processing step, the difference in training time between the two models was insignificant ($< 5$ minutes). 

The DTW scores did not observe the same major changes as the mean squared error rates, but this can be explained by the fact that DTW is a cumulative metric that continues to rise as the sequence length increases. When matching pairs of random sequences of 23 time steps vs. 30 time steps, the sequences with 30 time steps will have a greater DTW score on average. The fact that DTW remained roughly the same between the two trials indicates the interpolated model was able to closely track the MCIF despite the longer sequence length. 

Below are two graphs comparing predicted TACs from the interpolated and non-interpolated model. The MCIF curve is in red and the predicted curves based on the IDIF and myocardial tissues are shown in blue. Notice that the interpolated model is able to much more closely track the trail end of the curve without loosing any fidelity in mapping to the true MCIF at the peak. 

\begin{figure}[htbp]
  \centering
  \begin{subfigure}[b]{0.48\textwidth}
      \includegraphics[width=\textwidth]{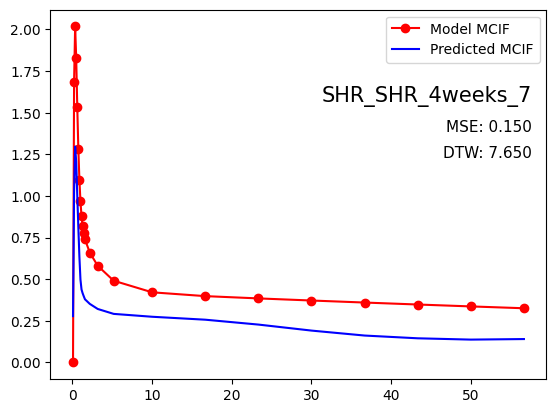}
      \caption{Non-interpolated LSTM output}
      \label{fig:no_interp}
  \end{subfigure}
  \hfill
  \begin{subfigure}[b]{0.48\textwidth}
      \includegraphics[width=\textwidth]{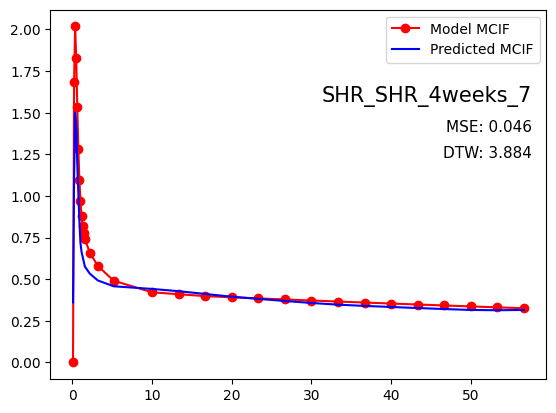}
      \caption{Interpolated LSTM output}
      \label{fig:interp}
  \end{subfigure}
  \caption{Comparison of model outputs with and without interpolation for a specific rodent. The x-axis is time in minutes and the y-axis is activity concentration.}
  \label{fig:lstm_comparison}
\end{figure}

Below are additional boxplots to show the spread of MSE and DTW scores across rodents in the interpolated datasets. 

\begin{figure}[H]
  \centering
  \begin{subfigure}[b]{0.48\textwidth}
      \includegraphics[width=\textwidth]{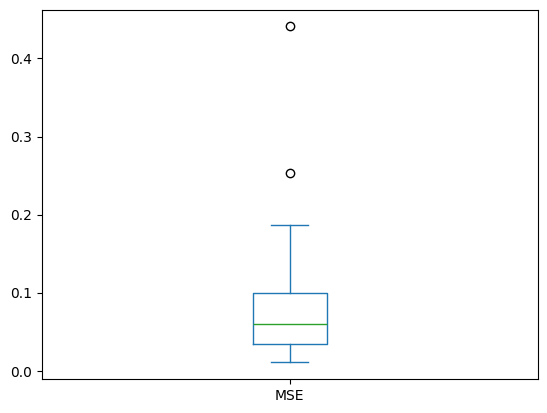}
      \caption{Boxplot for Mean Squared Error across all rodents}
      \label{fig:no_interp}
  \end{subfigure}
  \hfill
  \begin{subfigure}[b]{0.48\textwidth}
      \includegraphics[width=\textwidth]{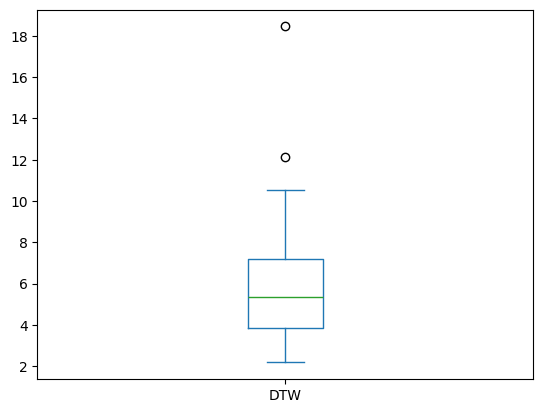}
      \caption{Boxplot for Dynamic Time Warping across all rodents}
      \label{fig:interp}
  \end{subfigure}
  \caption{Boxplots showing model spread across various metrics.}
  \label{fig:lstm_comparison}
\end{figure}

\section{Conclusion}
The results demonstrate that semi-automated segmentation of the IDIF and myocardium from dynamic PET data can effectively serve as inputs to an LSTM network for accurate MCIF prediction, even under constraints such as low sample size and heterogeneity in age, rodent type, and scanner configuration. This finding is significant given the importance of accurate MCIF estimation in non-invasive PET quantification.

Furthermore, interpolation was shown to mitigate data sparsity in the later phases of the TACs, improving prediction accuracy. This highlights the potential of simple preprocessing steps to enhance model performance in data-constrained biomedical contexts.

While promising, the study is limited by the dataset size and the use of semi-automated methods, which may introduce inconsistency. Future work will focus on increasing the dataset through additional scans and data augmentation, automating segmentation using U-Net architectures, and refining neural network models to better handle small, heterogeneous datasets.

To our knowledge, this is among the first demonstrations of data augmentation techniques in using deep sequential models trained on semi-automated dynamic PET data for MCIF prediction in preclinical rodent studies. These findings provide a foundation for future efforts toward fully automated and generalizable PET quantification pipelines.

\section{Acknowledgement}
This work was supported in part by a grant from the National Institutes of Health R01 HL123627 and start-up funds from the Department of Radiology and Medical Imaging at the University of Virginia (to BK). 

\section{References}
\bibliography{references} 

\end{document}